# Magnetic order and electronic transport properties in the Mn$_3$Al compound: the role of the structural state

V.V. Marchenkov*[a,b], V.Yu. Irkhin[a], E.B. Marchenkova[a], A.A. Semiannikova[a], P.S. Korenistov[a]

[a] M.N. Mikheev Institute of Metal Physics, 620077, S. Kovalevskaya str., 18, Ekaterinburg, Russia
[b] Ural Federal University, 620002, Mira str., 19, Ekaterinburg, Russia

**Abstract:** Electronic transport and magnetic properties of bulk and rapid melt quenched samples of the Mn$_3$Al Heusler alloy were studied. A correlation between the magnetic and structural states was established. For a cast sample, there is no ferromagnetic moment, and the behavior of the magnetic susceptibility (break at low temperatures and the Curie-Weiss law with high values of the paramagnetic Curie temperature) indicates a frustrated antiferromagnetic state. At the same time, for a rapid melt quenched sample, a ferrimagnetic state is observed with a moment close to compensation. The results of measurements of the electrical resistivity and the Hall effect evidence as well in favor of the implementation of these magnetic states.

*Vyacheslav Viktorovich Marchenkov, e-mail: march@imp.uran.ru, +7 (343) 374-53-83

**Introduction**

Many of the Heusler alloys [1] exhibit inconvenient magnetic and electronic properties [2–8]. In particular, states of an antiferromagnet and a compensated ferrimagnet, a topological semimetal and a frustrated magnet [7, 9–11] can be realized in them. These compounds have unusual characteristics, which are very sensitive to external influences and can be used in practice. The application of magnetic and electric fields, pressure, and temperature can allow "tuning" their electronic band structure, and, consequently, influence their transport and magnetic properties.

A special place among the large family of Heusler alloys is occupied by intermetallic compounds based on manganese $Mn_2Y Al$ ($Y$ = Cr, Fe, etc.) and $Mn_3Z$ ($Z$ = Al, Ga, Ge, Sn, etc.), which exhibit a number of interesting physical properties, like pure metallic manganese [12], including frustrated magnetic structures, a large anomalous Hall effect, and high values of electronic heat capacity [13–16].

The magnetic order of Heusler alloys exhibits a wide variety and is largely determined by the type of crystal structure. In Ref. [17], an almost zero magnetic moment was reported in thin films of the $Mn_3Al$ alloy, which was explained by the state of the compensated ferrimagnet. This state differs from antiferromagnetism by that the oppositely directed moments of manganese are located in different crystallographic positions, so that their possible compensation is not directly due to crystal symmetry, but is associated with a special band structure ("half-metallic antiferromagnetism") in the terminology of [9]).

It was suggested in [18] that the zero magnetization in cast $Mn_3Al$ alloy can be a manifestation of antiferromagnetism or compensated ferrimagnetism. The latter structure is very favorable for applications in spintronics.

**Materials and Methods**

It is well known that the structure of intermetallic compounds, in particular, $Mn_3Z$, can strongly depend on the methods of preparing alloys (cast, subjected to rapid melt quenching (RMQ) from the melt and thermobaric treatment, nano-structured, etc.), which inevitably reflects on their electronic and magnetic states (see, for example, [14, 17]). Regarding the magnetic structures of Heusler alloys, there are a number of calculated and experimental data, sometimes contradicting each other. In this work, basing on the study of the structure, electrical and magnetic properties of $Mn_3Al$, we will try to contribute to clarifying the picture for this compound.

The bulk polycrystalline $Mn_3Al$ alloy was synthesized in an induction furnace in a purified argon atmosphere. The prepared ingot was annealed for 72 h at 650°C in an argon atmosphere, followed by cooling to room temperature at a rate of 100 deg/h. RMQ $Mn_3Al$ tapes were obtained from a bulk compound by rapid quenching from a melt at room temperature under a pressure of 0.28 atm in a purified argon atmosphere. The RMQ tapes have a thickness of about 20 μm with a width of about 2 mm.

The certification of the casted alloys and the RMQ tapes was carried out by X-ray diffraction analysis and scanning electron microscopy at the Collaborative Access Center of the IMP UB RAS. The magneto-transport and magnetization were also measured at the Collaborative Access Center.

**Results and discussion**

Figure 1 shows the X-ray diffraction pattern of the $Mn_3Al$ compound in the initial cast state (a) and after rapid melt quenching (b). The bar chart (c) shows the position of the reflections corresponding to the structure of β-Mn. The reflections of both the cast and the RMQ alloy belong to the β-Mn structure, space group #213.

Figure 2 shows the temperature dependence of the electrical resistivity $\rho(T)$ of $Mn_3Al$. In the cast alloy, the residual resistivity $\rho_0$ is quite large and reaches a value of ~ 307 μΩ cm, $\rho(T)$ having a semiconductor form, i.e. decreases with temperature. A similar behavior with close values of $\rho$ was observed in thin films of Mn$_3$Al [17]. In terms of the resistivity value and the form of its temperature dependence, our data on the cast alloy are similar to [14], where Mn$_2$FeAl was studied and this behavior of $\rho(T)$ was explained by a large contribution of structural disorder.

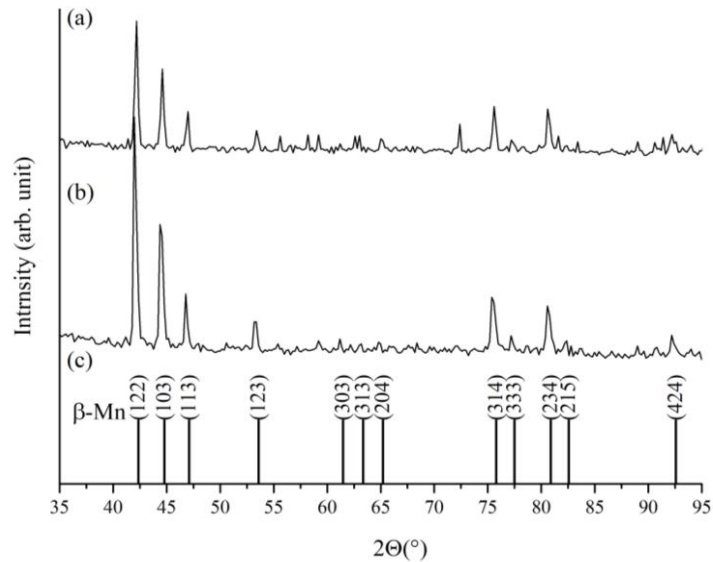

Fig. 1. X-ray diffraction pattern of the Mn$_3$Al alloy in the initial cast state (a), after rapid quenching from the melt (b), bar chart of β-Mn (c).

Quenching leads to great changes: a decrease in the residual electrical resistivity $\rho_0$ to 12.6 μΩ cm by more than an order of magnitude, the appearance of a minimum on the temperature dependence $\rho(T)$ at ~ 60 K, and a "metallic" behavior above 60 K (Fig. 2). The presence of the minimum on the ρ(T) curve and the negative temperature coefficient of resistivity (TCR) below 50 K, can be explained by the competition between various current carrier scattering mechanisms (see, for example, reviews [5, 7] and references therein).

The reasons for the TCR, i.e., an increase of resistivity with decreasing temperature, in metal alloys may be the following ones. There is so-called "Mooij rule" [19], according to which the metallic materials with structural disorder can have a negative TCR. Usually these are alloys with a high degree of disorder due to a plenty of grains and, accordingly, grain boundaries. The presence of structural disorder leads to negative TCR in Ref. [14].

Another possible reason for the appearance of a negative TCR and a minimum in the electrical resistivity is spin-dependent scattering on the interfaces (metal-metal, grain boundaries, etc.) [20]. In experimental work [21], a minimum in the electrical resistivity of manganites was observed, which was explained by the theory [20]. RMQ-alloy Mn$_3$Al contains grains and grain boundaries, and the charge-carrier scattering on them can lead to negative TCR. However, additional studies are required to unequivocally answer this question. Such studies are under the way.

It can be assumed that similar significant changes should also occur in the magnetic properties.

Figure 3 shows the field dependences of the magnetization $M$ for the cast and RMQ Mn$_3$Al alloy at $T = 4.2$ K. Obviously, in the case of the cast alloy, the magnetization $M$ is small in magnitude and increases weakly and linearly with the field: $M = 0.04$ μ$_B$/f.u. in the field 70 kOe.

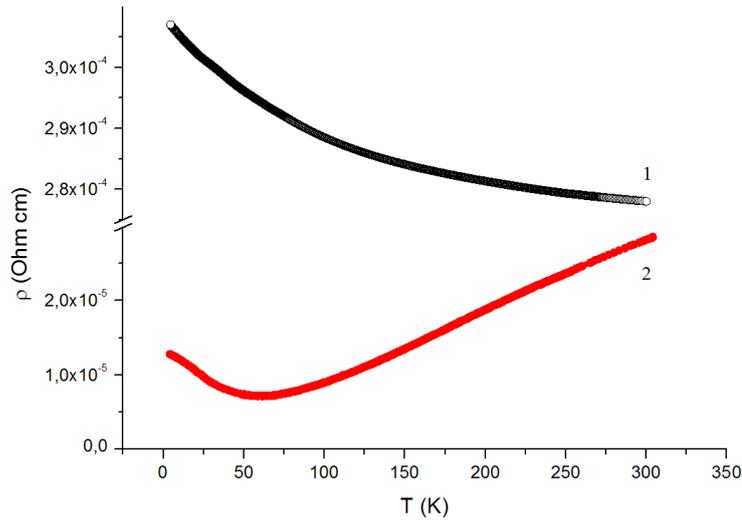

Fig. 2. Temperature dependences of the electrical resistivity of the Mn$_3$Al alloy: 1 – cast, 2 – RMQ.

In the case of the RMQ Mn$_3$Al alloy (Fig. 3), the shape of the *M*(*H*) drastically differs from the cast alloy. Already in low fields, a sharp increase in magnetization is observed. It is followed by a weaker increase and the absence of saturation, while in small fields hysteresis is observed. In this case, the magnetization is also small, and in a field of 70 kOe it is only 0.13 μ$_B$/f.u. In terms of the magnetization order of magnitude, our data are qualitatively similar to the data obtained on Mn$_3$Al films [17], which have the D0$_3$ structure according to [17]. The authors of this work estimated the magnitude of the magnetization as $M = 0.11 \pm 0.04$ μ$_B$/f.u. and concluded that the state of a compensated ferrimagnet was observed. It can be assumed that an almost compensated ferrimagnetic state is also realized in RMQ tapes.

Using the temperature dependences of the magnetization in the field (Fig. 4) for the cast Mn$_3$Al alloy, the magnetic susceptibility was determined, which strongly depends on temperature and approximately obeys the Curie-Weiss law with a magnetic moment of the order of the manganese ion moment and a paramagnetic Curie temperature θ$_{CW}$ ≈ - 300 K. This behavior indicates the realization of an antiferromagnetic state. The Neel temperature $T_N$ can be estimated from the magnetic susceptibility break at $T_N$ = 35 K, which is observed both in low (100 Oe) and high (50 kOe) fields (Fig. 4). The Neel temperature is determined similar to Refs. [14, 22], and antiferromagnetism of Mn$_3$Al is characterized by long-range order, since a more smooth maximum is expected in the case of a short-range order only. At the same time, low value of the Neel temperature in the frustrated system indicates the presence of considerable short-range order above this temperature [23].

In Ref. [22], where the magnetization of a cast Mn$_2$FeSi compound with the structure of an inverse Heusler alloy was studied, a behavior similar to ours was observed for the magnetic properties of the samples before and after quenching. The magnetization of the quenched alloy turned out to be small in magnitude and increased linearly with the magnetic field, and a break appeared on the temperature dependences in a field of 100 Oe, which was identified with the Neel temperature $T_N$ = 48 K. Quenching led to a slight increase in the magnetization value, its sharp increase in weak fields and more gradual increase in fields above 5 kOe without a tendency to saturation.

The *β*-Mn structure consists of two nonequivalent sublattices, one of which is a set of triangles located perpendicular to the directions of the set of axes [111] and forming a frustrated three-dimensional kagome-type lattice [12, 14]. Recent experimental studies have shown that in strongly frustrated (i.e., with competition of exchange interactions) systems, not only the quantum state of the spin liquid can arise, but also antiferromagnetism with a significantly reduced, but still finite Neel point. Such systems are characterized by the so-called frustration parameter, the ratio $|\theta_{CW}|/T_N$; in the intermediate temperature range $T_N < T < |\theta_{CW}|$ the system can exhibit unusual spin-liquid properties [23].

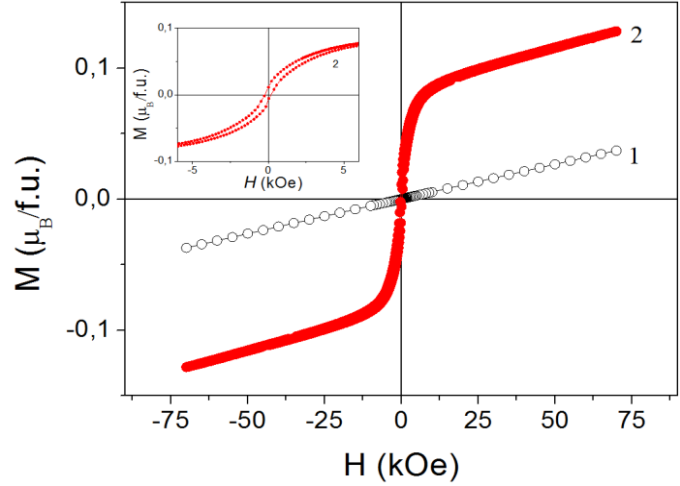

Fig. 3. Field dependences of the magnetization $M$ for the Mn$_3$Al alloy at $T = 4.2$ K 1 – cast, 2 – RMQ. The inset shows the hysteresis for the RMQ alloy.

High values of the frustration parameter are observed, for example, in the PdCrO$_2$ compound, where $T_N = 37$ K, $\theta_{CW} \approx -500$ K [24]. This behavior is apparently not described by the standard Heisenberg model and is due to correlation effects in the subsystem of itinerant electrons [25]. A similar behavior of the magnetic susceptibility for the Mn$_2$FeAl compound with the β-Mn structure was recently discovered and compared to the frustrated antiferromagnetism in [14, 15] ($T_N = 42$ K, $\theta_{CW} \approx -230$ K according to [15]).

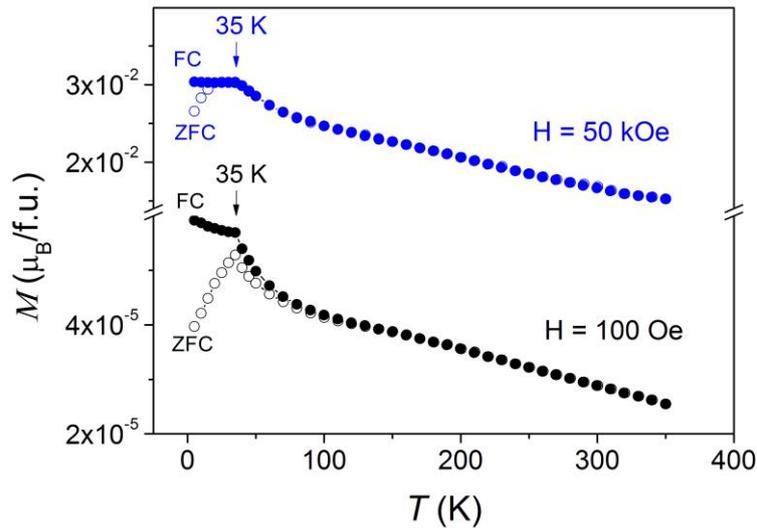

Fig. 4. Temperature dependence of the magnetization of cast Mn$_3$Al alloy in a magnetic field of 100 Oe and 50 kOe. Solid circles correspond to field cooling mode (FC), open circles to zero field cooling (ZFC).

In the case of the RMQ Mn$_3$Al ferrimagnetic alloy, an anomalous Hall effect should be observed. On the contrary, for a cast Mn$_3$Al alloy without a ferromagnetic moment, there should be no anomalous contribution. Figure 5 shows the field dependence of the Hall resistivity $\rho_{xy}$ at $T = 4.2$ K. In the cast $Mn_3Al$ alloy (Fig. 5, curve 1), a linear increase in $\rho_{xy}$ is observed, and there is no anomalous component.

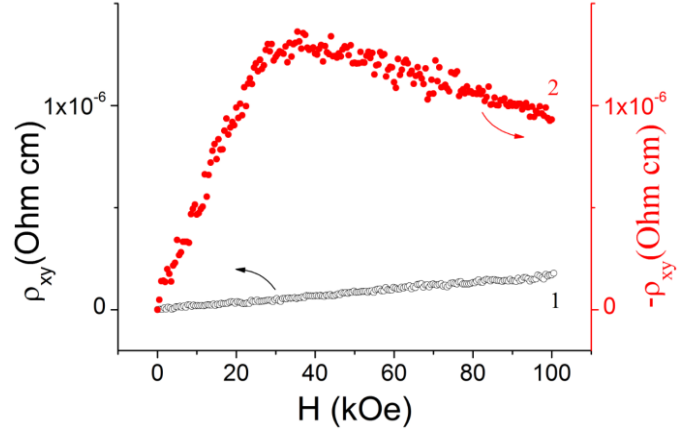

Fig. 5. Field dependences of the Hall resistivity $Mn_3Al$ alloy at $T = 4.2$ K: 1 – cast, 2 – RMQ.

In the case of an RMQ alloy (Fig. 5, curve 2), the behavior of $\rho_{xy}(H)$ is indeed characteristic of alloys with an anomalous Hall effect [26]. Results presented in Fig. 5 confirm our assumptions about the absence of spontaneous magnetization in the case of cast $Mn_3Al$ and the realization of the state of a ferrimagnet in the RMQ $Mn_3Al$ alloy.

Using the method described in [27], we determined the normal $R_0$ and anomalous $R_S$ Hall coefficients and estimated the concentration and mobility of current carriers. The results are presented in Table 1.

It can be seen that the anomalous Hall coefficient is equal to zero for the cast alloy, while for the RMQ tape the values of the normal and anomalous Hall coefficients differ by 3 orders of magnitude and have opposite signs, as is usually the case for magnets [26]. In Ref [28], a nontrivial anomalous Hall effect (AHE) in $Mn_3Al$ was studied, using symmetry arguments and first-principles calculations. It was shown that the explicit calculation of AHE also gives quite large Hall conductivity $\sigma_{xy} = -320$ (Ohm·cm)$^{-1}$ at the Fermi energy, which becomes as huge as $-1200$ (Ohm·cm)$^{-1}$ with hole doping. It should be noted the authors of [28] assumed the structure of $Mn_3Al$ in conventional cubic unit cell with space group #225, and the magnetic structure of the compensated ferrimagnet. In our case, the RMQ alloy has $\beta$-Mn structure, space group #213, and it is quite not a half-metallic ferromagnet. It should be noted that there is a huge difference in the concentration of current carriers, as well as in the mobility for the cast and RMQ alloys. As noted above, such a difference can be associated with a strong structural disorder in the cast alloy (see also [14]).

**Table 1.** Normal $R_0$ and anomalous $R_S$ Hall coefficients, concentration $n$, and mobility $\mu$ of current carriers of the cast and rapid melt quenched $Mn_3Al$ alloy.

| $Mn_3Al$ alloy | $R_0$, cm$^3$/C | $R_S$, cm$^3$/C | $n$, cm$^{-3}$ | $\mu$, cm$^2$/(s·V) |
|---|---|---|---|---|
| Cast | 1.78·10$^{-4}$ | - | 3.5·10$^{22}$ | 0.6 |
| RMQ | 1.83·10$^{-3}$ | -5.1 | 3.4·10$^{21}$ | 150 |

## Conclusions

Thus, the considered Mn$_3$Al Heusler alloy has a high sensitivity of the magnetic state and moment to its structural state. Therefore, the method of preparation and processing of the alloy plays a significant role in the formation of the electronic and magnetic characteristics of alloys under consideration.

According to calculations [29] for the Mn$_3$Al compound in the β-Mn structure, a ferrimagnet state is realized in which the magnetic moment of the sublattices is not significantly compensated. In this case, it follows from *ab initio* calculations [17, 30] in the D0$_3$ structure that a compensated ferrimagnetic state with a zero moment and a half-metallic structure arises. The results of [29] probably reflect some qualitative trend, but are not fully confirmed by our experimental data, according to which a frustrated antiferromagnetic and an almost compensated ferrimagnetic state, respectively, arise for both the cast and RMQ Mn$_3$Al alloys.

## Acknowledgements

The structure studies were carried out within the state assignment of Ministry of Science and Higher Education of the Russian Federation (themes "Structure" No. 122021000033-2 and "Spin" No. 122021000036-3).

Synthesis of the alloy and RMQ tapes, studies of electron transport and magnetic properties were supported by Russian Science Foundation (project No. 22-22-00935).

Authors thank P.B. Terentyev, D.A. Shishkin, V.N. Neverov for help and valuable discussions.